\documentclass[9pt,twocolumn,twoside]{osajnl}

\journal{ol} 

\setboolean{shortarticle}{true}

\title{Architecture for Integrated RF Photonic Downconversion of Electronic Signals}

\author[1,*]{Nathan P. O'Malley} 
\author[1]{Keith A. McKinzie}
\author[1,2]{Mohammed S. Alshaykh}
\author[3]{Junqiu Liu}
\author[1,4]{Daniel E. Leaird}
\author[3]{Tobias J. Kippenberg}
\author[1,5]{Jason D. McKinney}
\author[1]{Andrew M. Weiner}

\affil[1]{Elmore Family School of Electrical and Computer Engineering, Purdue University, West Lafayette, Indiana 47907, USA.} \affil[2]{Electrical Engineering Department, King Saud University, Riyadh 11421, Saudi Arabia.} 
\affil[3]{Institute of Physics, Swiss Federal Institute of Technology Lausanne (EPFL), CH-1015 Lausanne, Switzerland}
\affil[4]{Currently with Torch Technologies, supporting AFRL/RW, Eglin Air Force Base, Florida, USA}
\affil[5]{Formerly with U.S. Naval Research Laboratory, Washington, DC 20375, USA}

\affil[*]{Corresponding author: omalleyn@purdue.edu}

\begin{abstract}
Electronic analog to digital converters (ADCs) are running up against the well-known bit depth vs bandwidth tradeoff. Towards this end, RF photonic-enhanced ADCs have been the subject of interest for some time. Optical frequency comb technology has been used as a workhorse underlying many of these architectures. Unfortunately, such designs must generally grapple with SWaP concerns, as well as frequency ambiguity issues which threaten to obscure critical spectral information of detected RF signals. In this work, we address these concerns via an RF photonic downconverter with potential for easy integration and field deployment by leveraging a novel hybrid microcomb / electro-optic comb design.
\end{abstract}

\setboolean{displaycopyright}{false}

\begin{document} 

\maketitle

Increasing demand for high-speed data transmission and processing has helped motivate the development of electronic analog to digital converters (ADCs) at impressive speeds. Unfortunately, progress in ADCs seems to be challenged by the well-known bit-depth vs bandwidth tradeoff \cite{walden2008analog}. Technical issues with timing jitter often take a toll, and as transmission rates increase, so does comparator ambiguity \cite{walden2008analog}.  A range of RF photonic solutions have been proposed that vary in implementation \cite{khilo2012photonic, deakin2020dual, lukashchuk2019photonic, xie2012broadband, fang2021analog, ataie2015subnoise, li2018photonics}, but are generally united by their use of optical frequency combs. These combs possess extremely high frequency stability, and so promise to correct for some of the timing inaccuracies that plague electronic solutions at high frequencies. 

Demonstrations of RF photonic downconverters illustrate the power of optical frequency combs for sampling broadband electronic signals. However, in order to be deployed in real-world scenarios, optimization of size, weight, and power (SWaP) ought to be considered. Of the typical frequency comb generation methods, mode locked lasers are perhaps the most common, but cascaded electro-optic (EO) modulators are also frequently used. Both approaches can often be rather expensive from a SWaP perspective - commercially available mode locked lasers often have electrical-to-optical power conversion efficiencies of a few percent at most, with average optical powers at or above a few 100 mW \cite{diddams2010evolving}, suggesting electrical power draws of at least several Watts, often significantly more. Volumes of several liters are a typical lower bound. General SWaP metrics for EO comb generators are quite difficult to estimate due to the vast variety of implementations. However, as an example, a relatively typical EO comb generator commonly used in our lab requires powers and volumes roughly comparable with those of modelocked lasers \cite{metcalf2013high}.  This uses discrete optical components however, and significantly improved SWaP is possible with on-chip technologies such as lithium niobate on insulator (LNOI) modulators. A recent impressively low-power result (though having too few comb lines for the work presented in this Letter) still required more than 3 Watts of RF power \cite{ren2019integrated}. 

\begin{figure}
    \centering
    \includegraphics[width=8.8cm, height=2.9cm]{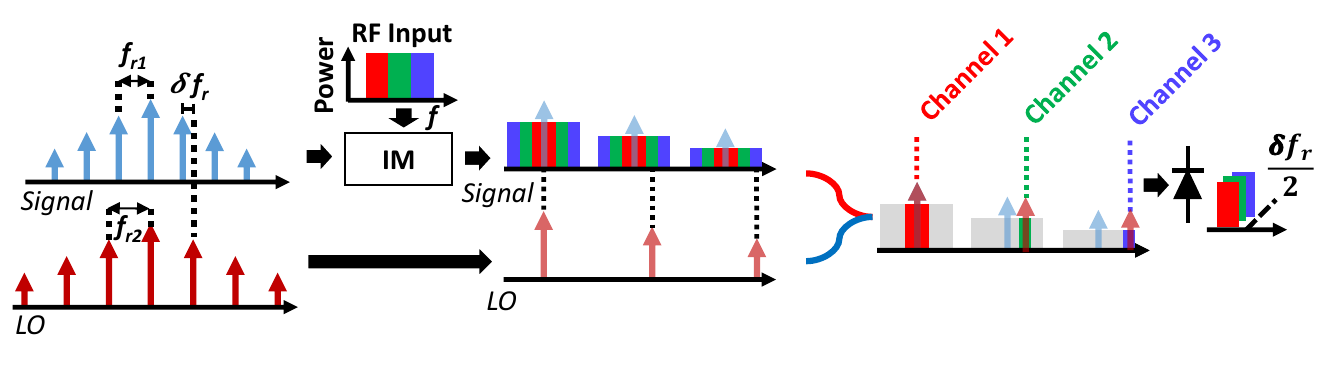}
    \caption{Diagram of dual comb approach for downconversion. LO comb lines each sample different portions of the RF input spectrum, thus folding the spectrum down to a $\delta f_r / 2$-wide baseband region.}
    \label{fig:figure1}
\end{figure}

Additionally, in many works on photonic downconversion, the photonic component of the system functions as a channelizing front end, splitting a broad input spectrum into slices and sending each slice to a separate electronic detector (see for example \cite{deakin2020dual}). While powerful, this approach adds complexity due to the need for a large number of parallel detectors. On the other hand, if the system is used to downconvert a broad input spectrum for a \emph{single} detector, it would suffer from ambiguity issues -- that is, while the electronic ADC stage may be able to digitize the signal of interest, spectral information is lost so the input signal's original frequency band is unknown after downconversion. This could pose significant problems for applications such as electronic support and signals intelligence \cite{adamy2001ew}. 

In this work, we demonstrate a dual-comb architecture for photonic downconversion of wideband RF signals with potential for eventual on-chip integration. We accomplish this by taking advantage of maturing silicon nitride microcomb technology for one of our two combs. Silicon nitride (SiN) resonators have been used to generate coherent soliton frequency combs on-chip for nearly a decade \cite{saha2013modelocking, herr2014temporal, wang2016intracavity, kippenberg2018dissipative}. The platform is CMOS foundry compatible and has been used to generate frequency combs with milliwatt levels of input optical power \cite{shen2020integrated}. Generally soliton microcombs require an off-chip pump source to drive the SiN microring. However, within the last few years, optical pump sources have been successfully integrated for soliton generation powered by a AAA battery \cite{stern2018battery} and even for heterogeneously integrated on-chip pump/microring pairs \cite{xiang2021laser}, suggesting convenient deployment of microcombs into the field may be approaching in the near future. 

While the SiN-based microcomb is very attractive from a SWaP perspective, the power per comb mode is limited by characteristically poor pump-to-sideband conversion efficiency \cite{bao2014nonlinear}. Thus, we utilize an electro-optic (EO) comb as our secondary comb, here termed the local oscillator (LO) comb. The EO comb can provide much higher power per comb mode than a soliton microcomb, which can lead to improved SNR. Additionally, the EO comb uniquely addresses another of the concerns with some previous RF downconverters -- namely, the issue of spectral ambiguity. Our hybrid EO comb/microcomb system - to our knowledge, a first-of-its-kind demonstration - allows for the use of a single detection stage (rather than many in parallel) to acquire the entire input spectrum while also overcoming the challenge of input frequency band ambiguity. This is accomplished through disambiguation techniques leveraging the easy tunability of the EO comb repetition rate.

The high-level view of our approach is outlined in figure \ref{fig:figure1}, and is related to works such as \cite{deakin2020dual, lukashchuk2019photonic, xie2012broadband}. The soliton ``signal'' comb has a repetition rate $f_{r1}$, and passes through a null-biased intensity modulator by which the RF signal of interest is inscribed onto the comb. Each comb mode inherits a spectral copy of the input RF spectrum. Note that since dual combs are used for downconversion, it is the mixing of the combs' electric field terms - rather than a single comb's intensity - that is of importance. Thus, the null bias point is appropriate rather than the quadrature bias often used in direct detection systems \cite{shieh2008coherent}. The LO comb has a repetition rate of $f_{r2} = f_{r1} + \delta f_r$. When the two combs beat together on a photodetector, the different LO comb modes essentially sample the input RF spectrum in chunks of bandwidth $\delta f_r$ (one so-called ``Nyquist zone'' of width $\delta f_r / 2$ on either side of each LO line) and fold them all down to a common baseband region from DC to $\delta f_r / 2$. This sampling of the RF spectrum is a result of the two combs' slightly offset repetition rates, and allows a system with bandwidth equal to that of the Nyquist zone width (rather than the full input spectral width) after the photonic downconverter to process the signal(s) of interest. 

\begin{figure}
    \centering
    \includegraphics[width=8.8cm, height=4.8cm]{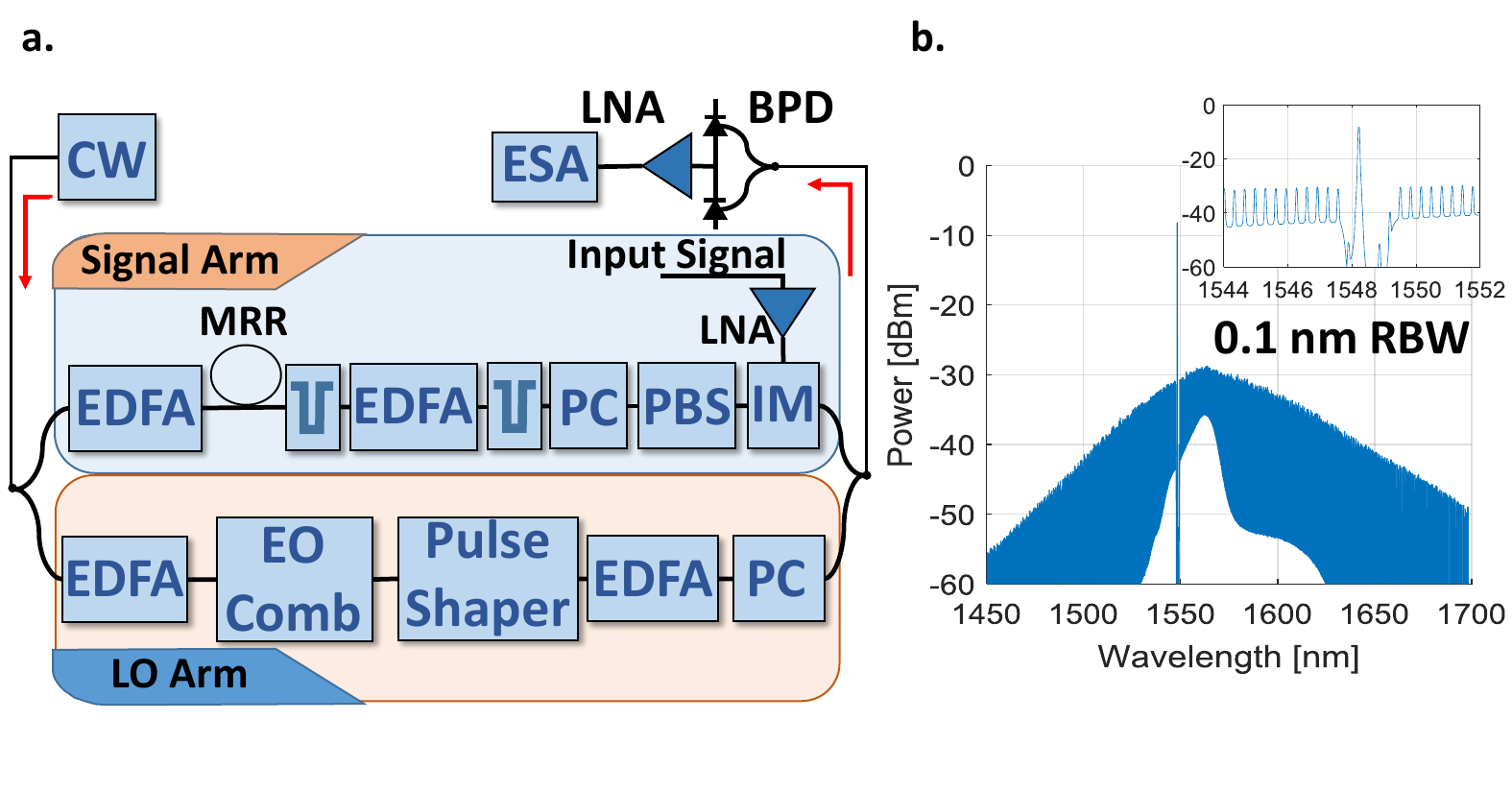}
    \caption{\textbf{a.} Experimental setup. CW - continuous wave laser; MRR - microring resonator; PC - polarization controller; PBS - polarizing beamsplitter; IM - intensity modulator; BPD - balanced photodetector; ESA - electrical spectrum analyzer. \textbf{b.} Optical spectrum of the soliton after the pump is partially notched by the filter after the ring. The EDFA's gain spectrum is visible from $\sim$1530-1620 nm. The inset is a close in of the comb near the pump, showing individual comb lines spaced at $\sim$40 GHz, as well as the spectral notching of the optical filter.}
    \label{fig:figure2}
\end{figure}

We implement this structure as in figure \ref{fig:figure2}a. An external cavity diode laser (ECDL) around 1550 nm is split into two arms, one for each comb. In the signal arm (soliton comb), an EDFA amplifies the laser before it is coupled on-chip via lensed fibers. Additionally, we rotate the polarization of the laser to properly align with the waveguides on chip. We utilize an ultrahigh-Q microring (FSR $\sim$ 40.374 GHz) from a batch featuring average loaded and intrinsic Q's of over 10$^7$ each, fabricated using the photonic Damascene process \cite{liu2021high}. This high Q lowers the required pump power for comb generation and allows for reduced thermal effects in the microring - effects that often plague soliton generation. Rather than using a complex generation scheme as is often required, we simply tune our laser from blue to red across a microring resonance until solitons are generated, and back-detune to reach a single soliton \cite{guo2017universal}. An OSA trace of the soliton's optical spectrum is shown in figure \ref{fig:figure2}b. The strong pump line (here attenuated by a DWDM functioning as a notch filter) is immediately visible above the weaker comb lines. A second EDFA and another DWDM-notch filter serve to increase the total comb power while reducing the relative pump power. The RF signal input is sent through an electrical low-noise amplifier (LNA) and modulated onto the signal comb using an optical intensity modulator (IM).

The LO arm (EO comb) begins similarly, in this case with an EDFA amplifying the ECDL before the optical carrier passes through the EO comb generator. This portion of our system has been explained elsewhere \cite{metcalf2013high}. In short however, a series of optical phase modulators and an intensity modulator are used to modulate the input CW laser, generating a flat-top, broadband optical frequency comb with 55 lines within its 3dB bandwidth. Due to the limited RF bandwidth of the modulators used to generate the EO comb, we use a tunable RF signal generator to set the comb’s repetition rate at $\sim$14.266 GHz, just over one third of the soliton repetition rate.  Accordingly, every EO third comb line is situated relatively close to a soliton line and can participate in downconversion.  This effectively multiplies the EO comb rep rate by three to 3 $\times$ 14.266 = 42.798 GHz. This is $\sim$2.42 GHz higher than the soliton repetition rate, fixing $\delta f_r$ = 2.42 GHz and the Nyquist bandwidth at 1.21 GHz.  A commercial pulse shaper is used to select desired lines from the EO comb.  For the experiments reported here, we only allow one LO line from the EO comb to pass at any given time, thus demodulating only two corresponding Nyquist bands.  Alternatively, the spectral filter could be configured to allow only every third EO comb line to pass, which would provide for downconversion of multiple RF bands simultaneously. The presence of multiple LO lines would increase shot noise, but due to large link loss (to be discussed shortly) our system is far from shot noise limited, so this is not expected to significantly degrade our system performance.

After the pulse shaper, an EDFA amplifies the remaining EO comb lines, which are then polarization-rotated to match the polarization of the signal comb before detection on a balanced photodetector. An LNA chain is used to resolve the system noise floor above our spectrum analyzer's instrument noise floor, but in a real-world application these would not be required, and the chain gain and noise figure (NF) have been removed from the results reported here.

\begin{figure}
    \centering
    \includegraphics[width=8.8cm, height=5cm]{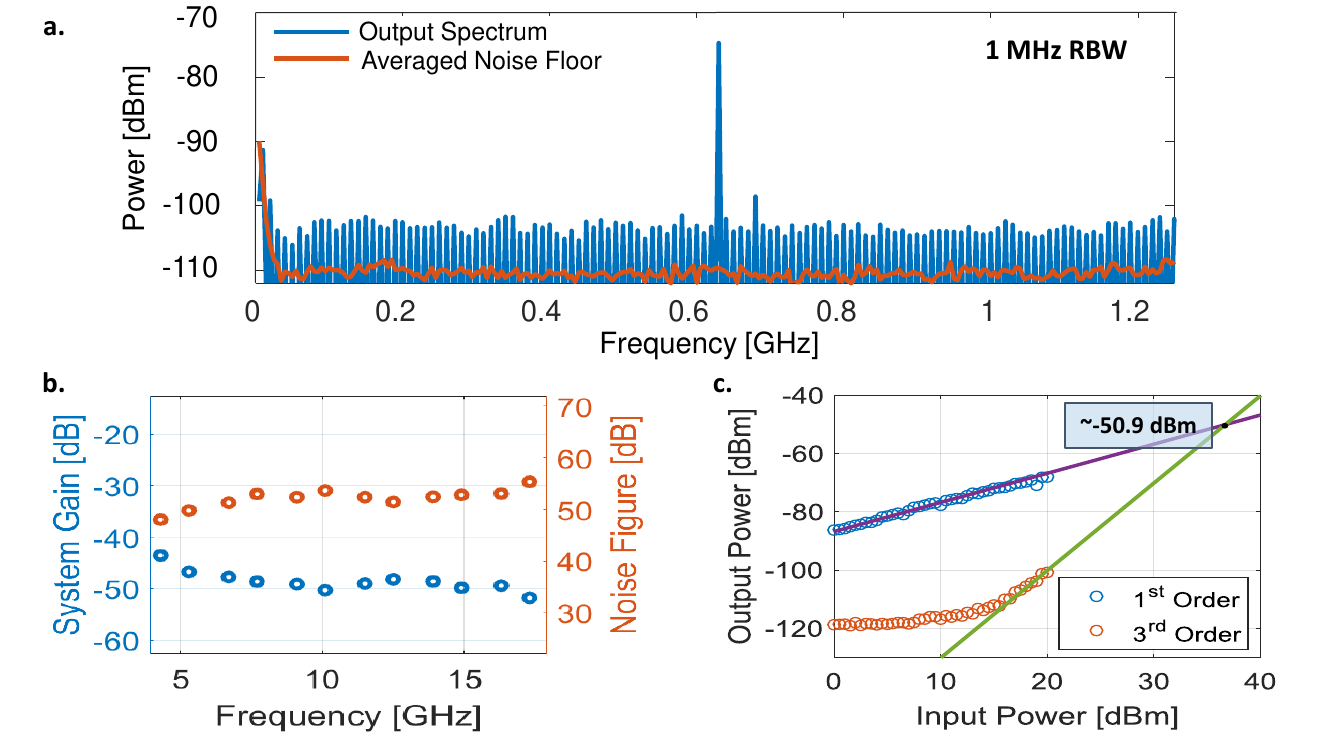}
    \caption{\textbf{a.} A sample system output RF spectrum. A small extraneous spur is seen around 675 MHz. The averaged system output noise floor is also shown. \textbf{b.} System gain and NF metrics. \textbf{c.} A two-tone test of the photonic portion of the system (without the pre-IM LNA) is conducted with RF inputs at $\sim$11.5 GHz.}
    \label{fig:figure3}
\end{figure}

Our system supports input RF signals from $\sim$3.6 GHz to $\sim$18 GHz, limited on the low side by our optical pump notch filters and on the high side by the bandwidth of our pre-IM LNA. We test our system performance by using single sinusoidal tones at various frequencies throughout the operating band. A summary of results can be seen in figures \ref{fig:figure3}a and \ref{fig:figure3}b. Figure \ref{fig:figure3}a is a trace taken from our ESA at 1 MHz resolution bandwidth (RBW) with an 11.5 GHz RF tone input to the system. It is downconverted to the expected $\sim$600 MHz, maintaining a $\sim$ 35 dB SNR at 1 MHz RBW with relatively limited RF input power around -30 dBm. A summary of metrics across the bandwidth of our system can be seen in figure \ref{fig:figure3}b. The photonic link's gain ranges from -44 dB to -51 dB over its operating bandwidth. The noise figure ranges from 48 dB to 55 dB. These gain/NF values are calculated by standard cascade analysis \cite{urick2015fundamentals} using measured gain and specified noise figure of the pre-IM LNA and measured gain and noise figure of the photonic link. The loss of the photonic portion of the system (roughly -80 dB) strongly dominates the gain of our pre-IM LNA (Gain $\sim$ 33 dB, noise figure $\sim$ 2.4 dB), resulting in a significant noise figure attributed predominantly to the system loss. The aforementioned low conversion efficiency of the microcomb is certainly a contributing factor here. 

After testing with single sinusoidal tones, we perform two-tone measurements of our pre-IM LNA and the rest of the photonic link separately in order to identify their individual output third-order intercept points (OIP$_3$), a standard metric for quantifying nonlinearity in RF systems \cite{pozar2005microwave}. Seen in figure \ref{fig:figure3}c is a measurement of the optical system with input tones around 11.5 GHz. The OIP$_3$ is measured to be roughly -50.9 dBm, determined from fixed-slope linear fits to the log-scale data as shown. The LNA OIP$_3$ is measured separately to be $\sim$18.4 dBm. A standard cascade analysis \cite{urick2015fundamentals} reveals a computed total system OIP$_3$ of $\sim$ -58 dBm with an 11.5 GHz input signal, again strongly limited by the photonic portion of the system, and again expected in a photonic system with relatively low comb powers. As a last step in analyzing the system's RF metrics, we use the measured noise and OIP$_3$ seen in figure \ref{fig:figure3}c to estimate the spur-free dynamic range (SFDR) of our system. We use the standard definition of SFDR$_3 = (OIP_3 / N_{out})^{2 / 3}$ \cite{pozar2005microwave}, and observe $N_{out}$ as $\sim$ -170 dBm / Hz  by examining the averaged noise floor in figure \ref{fig:figure3}a - thus, we estimate SFDR$_3 \approx 74.6$ dB / Hz$^{2/3}$. There are several promising options for further advancing the system performance in both linear and nonlinear metrics, which will be addressed in more detail shortly.

As all of the input spectrum's $\delta f_r / 2$-wide Nyquist zones are folded down into a single common $\delta f_r / 2$-wide band at the output, one cannot immediately discern a signal's originating Nyquist zone simply by detecting it at the output. This could be overcome by either limiting the allowed RF input frequencies (not desirable, as it would artificially limit the system bandwidth) or by using many parallel photodetector + ADC pairs after the downconverter (eg, \cite{fang2021analog}). In this work, we leverage the easy tunability of our EO comb repetition rate - as simple as turning a knob on the driving signal generator - to perform disambiguation \cite{harmon2014precision}. As the EO comb repetition rate is shifted very slightly, the output signal also shifts very slightly in frequency. Critically, the shift of the output signal is proportional to the product of the EO comb rep rate shift and the number of EO comb modes the LO line of interest is away from the original CW pump. Thus, simply by measuring the output signal frequency shift with a known rep rate shift, the originating Nyquist zone of the input signal can be immediately recovered. 

\begin{figure}
    \centering
    \includegraphics[width=8.8cm, height=5.0cm]{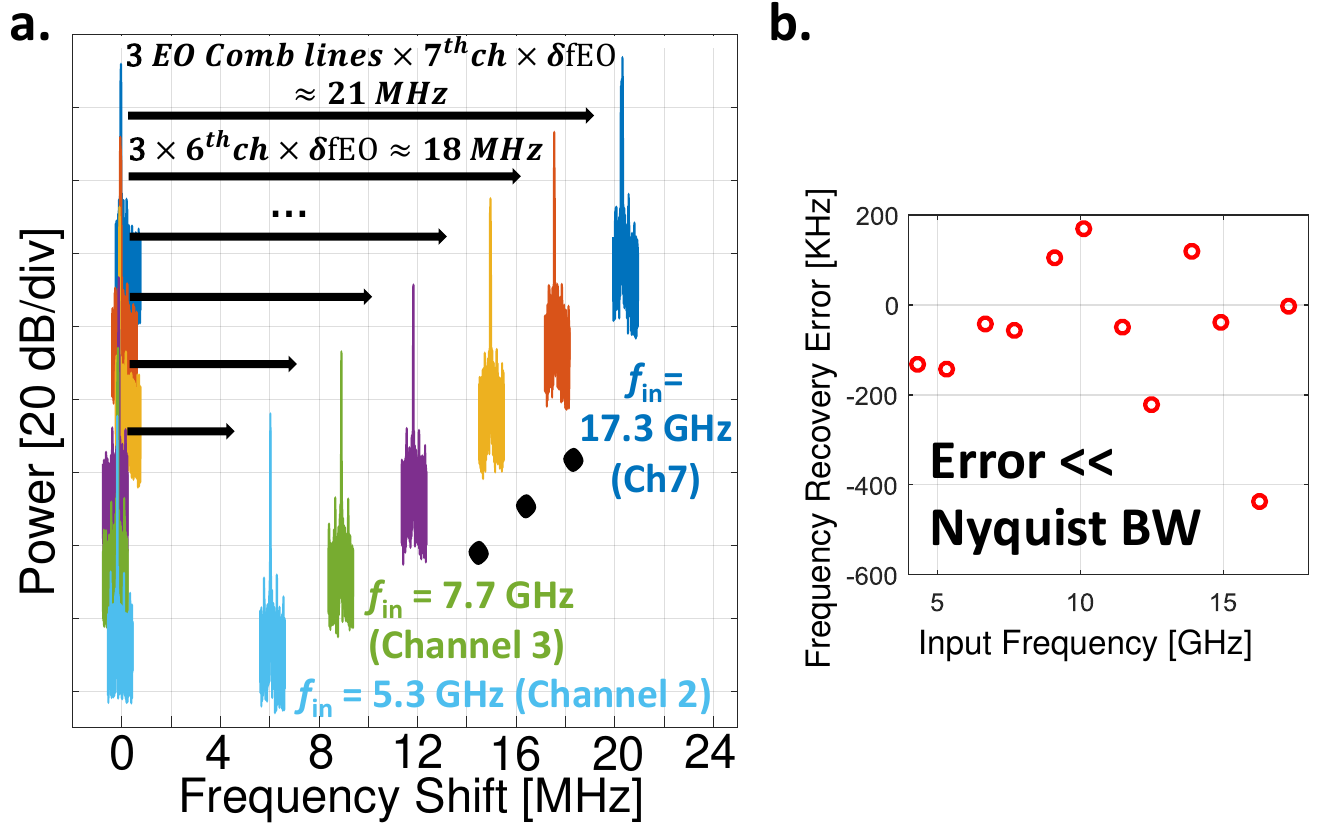}
    \caption{\textbf{a.} Demonstration of disambiguation. On the left, six test signals are separately downconverted to the baseband Nyquist zone. After the EO comb repetition rate is shifted by 1 MHz, the downconverted input at 5.3 GHz shifts by roughly 6 MHz while a downconverted 7.7 GHz signal shifts by 9 MHz, and so on. \textbf{b.} Recovered signal frequency error.}
    \label{fig:figure4}
\end{figure}

This technique is tested separately with signals at twelve different input Nyquist zones - six are ``low" Nyquist zones (Nyquist bands immediately at the low-frequency side of their respective LO line), and six are at ``high" Nyquist zones (bands at higher frequency than their LO line). Importantly, low and high Nyquist zones are distinguished by the direction of the output signal frequency shift when the EO comb rep rate is changed. This is a key benefit of our system, as this kind of Nyquist zone distinguishing is often accomplished via coherent detection (see for example, \cite{chen2014photonics}) but here requires only a single BPD. The results of this technique with the six high Nyquist bands are shown in figure \ref{fig:figure4}a. After disambiguation, the input frequencies are recovered; our results match the actual input frequencies (measured separately on a spectrum analyzer) with only a few hundred KHz of error, as seen in figure \ref{fig:figure4}b. We believe this error may be due to minor drifting in the free-running soliton repetition rate, but regardless is much smaller than the Nyquist bandwidth (1.21 GHz), allowing for unambiguous recovery. This disambiguation could be used to recover multiple signals simultaneously given a relatively sparse input spectrum.

Finally, we believe there are substantial improvements that can still be made. One simple tactic would be adding a broader bandwidth, higher-gain LNA before our IM. Our current LNA has a gain of roughly 33 dB which could be dramatically improved with an appropriate replacement. Additionally, upgrading our EO comb generator with higher bandwidth modulators (>40 GHz) could allow for removal of the pulse shaper, reducing optical insertion loss in the system and boosting our SNR. Finally, it has been noted that the conversion efficiency of the soliton microcomb appears to be costing significantly in terms of performance. There has recently been exciting work in a particular brand of microcombs operating in a normal-dispersion regime (different from the anomalous dispersion regime used in this work) that boasts efficiencies one or even two orders of magnitude better than the microcombs here \cite{xue2015microresonator, kim2019turn}.  These so-called ``dark pulses'' typically have a comparatively narrow bandwidth, but have nevertheless already been demonstrated at repetition rates similar to ours while roughly spanning the spectral width used in this work \cite{jin2021hertz}. Replacing our microcomb with a normal-GVD comb could thus be a relatively simple adjustment resulting in significant improvement of our system.

In summary, we have demonstrated a dual comb RF photonic system for downconversion of signals in the 3.6-18 GHz band. It can potentially serve as a front end for low-bandwidth, high-resolution electronics. Additionally, we believe there is substantial potential for moving our system towards a fully on-chip downconverter. Recent advances in fully integrated soliton microcombs have already been highlighted here. While our EO comb in this case uses discrete optical components, advances in thin film lithium niobate modulators (e.g., \cite{hu2021high}) suggest our EO comb generator could be integrated in the near future. Furthermore, in addition to possessing potential for a deployable integrated system, our approach avoids the need for many parallel photodetectors, requiring only a single balanced pair. We believe the relative simplicity of such a system makes it a promising low-SWaP concept for applications such as mobile signals intelligence or even multi-band 5G backhaul systems (e.g., \cite{zhang2014low}). 

\begin{backmatter}
\bmsection{Funding}
National Science Foundation 1809784-ECCS, US Naval Research Laboratory N00173-18-S-BA01, Air Force Office of Scientific Research FA9550-19-1-0250, Swiss National Science Foundation 176563 (BRIDGE), The College of Engineering Research Center at KSU, National Defense Science and Engineering Graduate Fellowship Program

\bmsection{Acknowledgments}
Thanks to Anton Lukashchuk for helpful comments on the manuscript. Portions of this work were presented at CLEO 2022, paper number STh5M.7.

\bmsection{Disclosures}
The authors declare no conflicts of interest.

\end{backmatter}

\section{Extended Bibliography for Reference}

\begin{enumerate}
\item Walden, R.H., 2008. Analog‐to‐Digital Conversion in the Early Twenty‐First Century. Wiley Encyclopedia of Computer Science and Engineering, pp.126-138.

\item Khilo, A., Spector, S.J., Grein, M.E., Nejadmalayeri, A.H., Holzwarth, C.W., Sander, M.Y., Dahlem, M.S., Peng, M.Y., Geis, M.W., DiLello, N.A. and Yoon, J.U., 2012. Photonic ADC: overcoming the bottleneck of electronic jitter. Optics Express, 20(4), pp.4454-4469.

\item Deakin, C. and Liu, Z., 2020. Dual frequency comb assisted analog-to-digital conversion. Optics Letters, 45(1), pp.173-176.

\item Lukashchuk, A., Riemensberger, J., Liu, J., Marin-Palomo, P., Karpov, M., Koos, C., Bouchand, R. and Kippenberg, T.J., 2019. Photonic-assisted analog-to-digital conversion using integrated soliton microcombs. In 45th European Conference on Optical Communication (ECOC 2019) (pp. 1-4). IET.

\item Xie, X., Dai, Y., Xu, K., Niu, J., Wang, R., Yan, L. and Lin, J., 2012. Broadband photonic RF channelization based on coherent optical frequency combs and I/Q demodulators. IEEE Photonics Journal, 4(4), pp.1196-1202.

\item Fang, D., Drayß, D., Lihachev, G., Marin-Palomo, P., Peng, H., Füllner, C., Kuzmin, A., Liu, J., Wang, R., Snigirev, V. and Lukashchuk, A., 2021, September. 320 GHz analog-to-digital converter exploiting Kerr soliton combs and photonic-electronic spectral stitching. In 2021 European Conference on Optical Communication (ECOC) (pp. 1-4). IEEE.

\item Ataie, V., Esman, D., Kuo, B.P., Alic, N. and Radic, S., 2015. Subnoise detection of a fast random event. Science, 350(6266), pp.1343-1346.

\item Li, Y., Kuse, N. and Fermann, M., 2018. Photonics-enabled wideband microwave burst detection. Optics Letters, 43(7), pp.1491-1494.

\item Diddams, S.A., 2010. The evolving optical frequency comb. JOSA B, 27(11), pp.B51-B62.

\item Metcalf, A.J., Torres-Company, V., Leaird, D.E. and Weiner, A.M., 2013. High-power broadly tunable electrooptic frequency comb generator. IEEE Journal of Selected Topics in Quantum Electronics, 19(6), pp.231-236.

\item Ren, T., Zhang, M., Wang, C., Shao, L., Reimer, C., Zhang, Y., King, O., Esman, R., Cullen, T. and Lončar, M., 2019. An integrated low-voltage broadband lithium niobate phase modulator. IEEE Photonics Technology Letters, 31(11), pp.889-892.

\item Adamy, D., 2001. EW 101: A first course in electronic warfare (Vol. 101). Artech house.

\item Saha, K., Okawachi, Y., Shim, B., Levy, J.S., Salem, R., Johnson, A.R., Foster, M.A., Lamont, M.R., Lipson, M. and Gaeta, A.L., 2013. Modelocking and femtosecond pulse generation in chip-based frequency combs. Optics express, 21(1), pp.1335-1343.

\item Herr, T., Brasch, V., Jost, J.D., Wang, C.Y., Kondratiev, N.M., Gorodetsky, M.L. and Kippenberg, T.J., 2014. Temporal solitons in optical microresonators. Nature Photonics, 8(2), pp.145-152.

\item Wang, P.H., Jaramillo-Villegas, J.A., Xuan, Y., Xue, X., Bao, C., Leaird, D.E., Qi, M. and Weiner, A.M., 2016. Intracavity characterization of micro-comb generation in the single-soliton regime. Optics express, 24(10), pp.10890-10897.

\item Kippenberg, T.J., Gaeta, A.L., Lipson, M. and Gorodetsky, M.L., 2018. Dissipative Kerr solitons in optical microresonators. Science, 361(6402), p.eaan8083.

\item Shen, B., Chang, L., Liu, J., Wang, H., Yang, Q.F., Xiang, C., Wang, R.N., He, J., Liu, T., Xie, W. and Guo, J., 2020. Integrated turnkey soliton microcombs. Nature, 582(7812), pp.365-369.

\item Stern, B., Ji, X., Okawachi, Y., Gaeta, A.L. and Lipson, M., 2018. Battery-operated integrated frequency comb generator. Nature, 562(7727), pp.401-405.

\item Xiang, C., Liu, J., Guo, J., Chang, L., Wang, R.N., Weng, W., Peters, J., Xie, W., Zhang, Z., Riemensberger, J. and Selvidge, J., 2021. Laser soliton microcombs heterogeneously integrated on silicon. Science, 373(6550), pp.99-103.

\item Bao, C., Zhang, L., Matsko, A., Yan, Y., Zhao, Z., Xie, G., Agarwal, A.M., Kimerling, L.C., Michel, J., Maleki, L. and Willner, A.E., 2014. Nonlinear conversion efficiency in Kerr frequency comb generation. Optics letters, 39(21), pp.6126-6129.

\item Shieh, W., Bao, H. and Tang, Y., 2008. Coherent optical OFDM: theory and design. Optics express, 16(2), pp.841-859.

\item Liu, J., Huang, G., Wang, R.N., He, J., Raja, A.S., Liu, T., Engelsen, N.J. and Kippenberg, T.J., 2021. High-yield, wafer-scale fabrication of ultralow-loss, dispersion-engineered silicon nitride photonic circuits. Nature communications, 12(1), pp.1-9.

\item Guo, H., Karpov, M., Lucas, E., Kordts, A., Pfeiffer, M.H., Brasch, V., Lihachev, G., Lobanov, V.E., Gorodetsky, M.L. and Kippenberg, T.J., 2017. Universal dynamics and deterministic switching of dissipative Kerr solitons in optical microresonators. Nature Physics, 13(1), pp.94-102.

\item Urick, V.J., Williams, K.J. and McKinney, J.D., 2015. Fundamentals of microwave photonics. John Wiley \& Sons.

\item Pozar, D.M., 2005. Microwave engineering. John wiley \& sons.

\item Harmon, S.R. and Mckinney, J.D., 2014. Precision broadband RF signal recovery in subsampled analog optical links. IEEE Photonics Technology Letters, 27(6), pp.620-623.

\item Chen, H., Li, R., Lei, C., Yu, Y., Chen, M., Yang, S. and Xie, S., 2014. Photonics-assisted serial channelized radio-frequency measurement system with Nyquist-bandwidth detection. IEEE Photonics Journal, 6(6), pp.1-7.

\item Xue, X., Wang, P.H., Xuan, Y., Qi, M. and Weiner, A.M., 2017. Microresonator Kerr frequency combs with high conversion efficiency. Laser \& Photonics Reviews, 11(1), p.1600276.

\item Kim, B.Y., Okawachi, Y., Jang, J.K., Yu, M., Ji, X., Zhao, Y., Joshi, C., Lipson, M. and Gaeta, A.L., 2019. Turn-key, high-efficiency Kerr comb source. Optics letters, 44(18), pp.4475-4478.

\item Jin, W., Yang, Q.F., Chang, L., Shen, B., Wang, H., Leal, M.A., Wu, L., Gao, M., Feshali, A., Paniccia, M. and Vahala, K.J., 2021. Hertz-linewidth semiconductor lasers using CMOS-ready ultra-high-Q microresonators. Nature Photonics, 15(5), pp.346-353.

\item Hu, Y., Yu, M., Buscaino, B., Sinclair, N., Zhu, D., Shams-Ansari, A., Shao, L., Zhang, M., Kahn, J.M. and Loncar, M., 2021, May. High-efficiency and broadband electro-optic frequency combs using coupled lithium-niobate microresonators. In CLEO: Science and Innovations (pp. STu2G-2). Optica Publishing Group.

\item Zhang, J.A., Ni, W., Matthews, J., Sung, C.K., Huang, X., Suzuki, H. and Collings, I., 2014, June. Low latency integrated point-to-multipoint and E-band point-to-point backhaul for mobile small cells. In 2014 IEEE international conference on communications workshops (ICC) (pp. 592-597). IEEE.

\end{enumerate}

© 2023 Optica Publishing Group. Users may use, reuse, and build upon the article, or use the article for text or data mining, so long as such uses are for non-commercial purposes and appropriate attribution is maintained. All other rights are reserved.





\end{document}